\documentclass[floatfix,footinbib,reprint,twocolumn,prx,aps,a4paper,preprintnumbers,amsmath,amssymb,longbibliography,superscriptaddress,nobibnotes]{revtex4-1}
\usepackage{graphicx}
\usepackage{subfigure}
\usepackage{enumitem}  
\usepackage{bbold} 
\DeclareGraphicsExtensions{.png,.pdf}
\usepackage{epsfig}
\usepackage{bm}
\usepackage{color}
\usepackage{makeidx}
\usepackage[colorlinks=true,linkcolor=blue,citecolor=blue,urlcolor=blue]{hyperref}
\usepackage{siunitx}
\usepackage{xcolor}
\usepackage{soul}
\setstcolor{red}
\usepackage{makecell}

\usepackage[bbgreekl]{mathbbol}
\DeclareSymbolFontAlphabet{\amsmathbb}{AMSb}%
\DeclareSymbolFontAlphabet{\mathbb}{AMSb}
\usepackage{amsfonts}
\usepackage{mathbbol}
\usepackage[utf8]{inputenc}
\usepackage[USenglish]{babel}

\begin{document}
\title{\bf
Fully Compensated Lines in Ferrimagnets
}

\author{Qais Ali}
\email[Contact author: ]{ali.qais@donau-uni.ac.at}
\affiliation{Department of Computational Materials Design, Max-Planck-Institut for Sustainable Materials, 40237 Düsseldorf, Germany}
\affiliation{Christian Doppler Laboratory for magnet design through physics informed machine learning, Department for Integrated
Sensor Systems, University for Continuing Education Krems, Viktor Kaplan-Stra\ss e 2E, 2700 Wiener Neustadt, Austria}
\author{Anna Gr\"{u}nebohm}
\affiliation{Interdisciplinary Center for Advanced Materials Simulation, Center for Interface-Dominated High
Performance Materials (ZGH) and Faculty for Physics and Astronomy, Ruhr-Universität Bochum, Universitätsstraße 150, 44801 Bochum, Germany}
\author{Halil \ifmmode \dot{I}\else \.{I}\fi{}brahim S\"{o}zen}
\affiliation{Institute of Chemistry, Carl-von-Ossietzky University of Oldenburg, Oldenburg D-26129, Germany}
\affiliation{Mercedes-Benz Turk A.S., Orhan Gazi Mahallesi, Mercedes Bulvari No. 17/1, 34519 Esenyurt, Istanbul, Turkey}
\author{Tilmann Hickel}
\affiliation{Department of Computational Materials Design, Max-Planck-Institut for Sustainable Materials, 40237 Düsseldorf, Germany}
\affiliation{BAM Federal Institute for Materials Research and Testing, 12489 Berlin, Germany}
\author{Jörg Neugebauer}
\affiliation{Department of Computational Materials Design, Max-Planck-Institut for Sustainable Materials, 40237 Düsseldorf, Germany}
\author{Eduardo Mendive-Tapia}
\email[Contact author: ]{e.mendive.tapia@ub.edu}
\affiliation{Departament de Física de la Matèria Condensada, Facultat de Física, Universitat de Barcelona, 08028 Barcelona, Catalonia}
\affiliation{Department of Computational Materials Design, Max-Planck-Institut for Sustainable Materials, 40237 Düsseldorf, Germany}
\affiliation{Institut de Nanociència i Nanotecnologia (IN2UB), Universitat de Barcelona, 08028 Barcelona, Catalonia, Spain}
\affiliation{Institut de Química Teòrica i Computacional (IQTCUB), Universitat de Barcelona, 08028 Barcelona, Catalonia, Spain}

\begin{abstract}
We generalize the classic Néel diagram and identify another type of ferrimagnetic phase that remains fully magnetically compensated below the Curie temperature, forming a continuous line of compensated points.
It exhibits zero net magnetization while retaining non-relativistic, eV-scale reciprocal spin splitting. We further find a persistently enhanced intrinsic switching field over a broad temperature range near the fully compensated phase. Proximity to the compensated line is achieved by minimizing the net local moment while balancing exchange interactions with respect to the number of equivalent atoms in each sublattice. The resulting extended Néel diagram provides practical design rules for engineering fully compensated ferrimagnetic phases via targeted chemical substitution that combines atoms with robust and weak local moments, as demonstrated through density functional theory and Monte Carlo simulations for GdCo$_5$-type ferrimagnets.

\end{abstract}

\maketitle

\section{Introduction}

Néel’s 1948 work laid the foundation for the theory of ferrimagnets~\cite{neel_pioneer_1948}. Smart later refined this framework by developing the well-known Néel diagram, which remains central for interpreting characteristic ferrimagnetic magnetization curves~\cite{Smart}. Ferrimagnets share features of both ferromagnets and antiferromagnets: below the transition temperature from the paramagnetic state $T_\text{tr}$, they exhibit spontaneous magnetization, yet neighboring ions are antiferromagnetically coupled, yielding two oppositely aligned sublattices~\cite{spaldin_2010}. The temperature dependence of this spontaneous magnetization can take forms far more varied than in simple ferromagnets, governed by the magnetic ions in each sublattice and by the molecular-field coefficients~\cite{Smart}.

\begin{figure}[t]
    \centering
    \includegraphics[scale=.72]{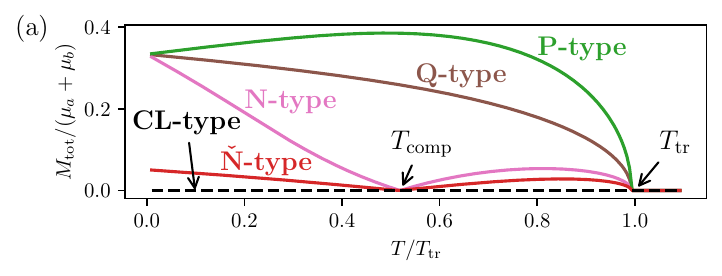}
    \includegraphics[scale=.40]{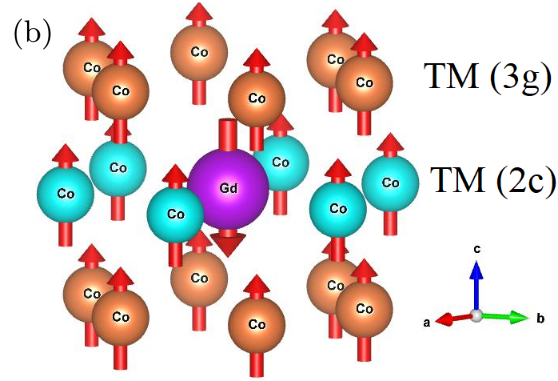}
    \caption{(a) Temperature dependency of the total magnetization, divided by the addition of the two sublattice moment sizes, for P (green), Q (brown) and N-type (pink) typical ferrimagnets with $\mu_a^\text{tot}=2\mu_b^\text{tot}$.
    A CL-type (fully compensated phase, dashed black line) as well as an intermediate \v{N}-type (low total magnetization, red line) are also shown.
    (b) CaCu$_5$ hexagonal structure of (Gd,TM$_{5}$)-type compounds with a ferrimagnetic ground state. In the unit cell there is one Gd site in purple and two different Wyckoff positions for TM atoms, namely $2c$ in cyan and $3g$ in golden brown. These TM sites can be occupied by one or a solution of different TM atoms (e.g., Co, Ni, Cu, Fe).}
    \label{fig:intro}
\end{figure}

The Néel diagram classifies ferrimagnets into three types: P, Q, and N~\cite{Smart}; see Fig.~\ref{fig:intro}(a). Q-type ferrimagnets resemble ferromagnets, showing a monotonic decrease of magnetization with temperature.
In P- and N-type ferrimagnets, however, the sublattices disorder at very different rates. In P-type systems, the sublattice with the smaller zero-temperature moment decays faster, producing a maximum in the total net magnetization at finite temperature~\cite{Neel,chris}. In N-type systems, the larger-moment sublattice disorders more rapidly, giving rise to a compensation point at temperature $T_\text{comp}$ where the total net magnetization vanishes.

The compensation point in ferrimagnets enables a wide range of applications, with the nearby angular-momentum compensation point playing a key role in the emergence of ultrafast dynamics. Their tunability through chemical composition makes these materials especially promising. Relevant technologies include 
ultrafast and low-power spintronic devices~\cite{sala2022ferrimagnetic, LIU202515_ultrafast, Zhang2023_FiMs_apps}, all-optical magnetic switching~\cite{kryder1993magneto, all_optical_PRL_Stanciu}, high-sensitivity magnetic sensors~\cite{Nora_GdCoCu}, and magnonics~\cite{kim2020distinct}. The temperature dependence of intrinsic material properties near the compensation point is an important and active area of research~\cite{chris_julie,K_A_PhysRevB.111.184416}.

In this work we develop an extended Néel’s classical diagram, shown in Fig.~\ref{fig:diagram1},  which reveals a previously unreported fully compensated phase where the magnetic-moment compensation remains complete over a broad temperature range below the transition temperature from the paramagnetic state, $T_\text{tr}$. This phase forms a continuous line of fully compensated points with zero net magnetization, akin to antiferromagnets, which we label as CL-type (Compensation-Line type), as shown in Fig.\ \ref{fig:intro}(a). Moreover, it provides a framework for understanding and, through chemical doping, guiding the realization of two key phenomena that arise at the full compensation and are important for technological applications.

The first concerns the pronounced increase in coercivity at $T_\text{comp}$ in typical N-type ferrimagnets, which serves as a practical marker for approaching the temperature range in which ferrimagnetic dynamics cross over to an ultrafast~\cite{PhysRevB.73.220402}, antiferromagnet-like regime. Prior work has shown that the coercive field peaks at $T_\text{comp}$ in rare-earth alloys~\cite{amy,Ostler}. However, how such an enhanced coercivity can be optimized and extended in temperature through modifications of material parameters remains unclear.

Second, we show how the CL-type ferrimagnetic phase exhibits zero net magnetization together with strong non-relativistic spin splitting in the reciprocal space. Although the microscopic origin differs from symmetry-driven altermagnetism, such fully compensated lines may display related spin-dependent transport and dynamical phenomena associated with spin-split electronic structures~\cite{doi:10.1073/pnas.2108924118,PhysRevX.12.040501,PhysRevX.12.040002,PhysRevX.12.031042,https://doi.org/10.1002/adfm.202409327,npjCheong}: ultrafast magnetic dynamics and strong (eV-scale) spin-exchange splitting, which underlies phenomena related to spin-splitting  currents~\cite{PhysRevLett.126.127701,PhysRevLett.133.056701} and torques~\cite{PhysRevB.102.014422,PhysRevB.102.144441,PhysRevB.103.125114,PhysRevLett.126.127701,PhysRevLett.128.197202,PhysRevLett.129.137201,PhysRevLett.134.176401,Zhang2025}, Hall effects~\cite{Smejkal2022,PhysRevLett.130.036702,Reichlova2024}, and giant or tunneling magnetoresistance~\cite{PhysRevX.12.011028}.

The ferrimagnetic systems considered in this work are rare-earth (RE)–transition-metal (TM) GdCo$_{5}$-type compounds. These materials crystallize in the hexagonal CaCu$_5$-type structure (space group $P6/mmm$). Along the $c$ axis, layers of TM atoms at the $3g$ Wyckoff site alternate with mixed layers containing Gd and TM atoms at the $2c$ Wyckoff site, see Fig.\ \ref{fig:intro}(b). Investigation of pseudo-binary series, such as Gd(Co$_{5-x}$Ni$_x$)~\cite{chuang}, reveals a rich compositional dependence in their magnetic behavior, including the change of $T_ \text{tr}$, $T_ \text{comp}$, and coercive field~\cite{amy,Ostler,RADWANSKI198657,Stinschoff_PhysRevB.95.060410}.

The paper is organized as follows. In Section~\ref{theory}, we first introduce the theoretical framework employed in this work. We begin by presenting a mean-field model for describing the temperature-dependent properties of ferrimagnets, which is subsequently applied to GdCo$_5$-type systems. The magnetic interactions entering the model are evaluated using first-principles calculations and further validated through complementary Monte Carlo simulations.
The results are presented in Section~\ref{results}. We discuss the extended Néel diagram and the \textit{ab initio}-informed composition--temperature phase diagrams, together with the non-relativistic exchange spin splitting and the temperature dependence of the coercive field in the fully compensated phase.

\section{Theoretical framework}
\label{theory}

\subsection{Model}

\subsubsection{Magnetic Hamiltonian and zero-temperature limit}
We model a uniaxial ferrimagnetic material with the following Heisenberg-type Hamiltonian
\begin{equation}
\mathcal{H} = -\frac{1}{2}\sum_{ij}J_{ij}\hat{\textbf{e}}_i\cdot\hat{\textbf{e}}_j
+\sum_i K_i\left[1- (\hat{\textbf{e}}_i \cdot \hat{\textbf{u}})^2\right]
-\textbf{B}\cdot\sum_i\mu_i\hat{\textbf{e}}_i
,
\label{eq_ham_real}
\end{equation}
where the factor $\frac{1}{2}$ removes the double counting, $\textbf{B}$ is the applied magnetic field, and $\{\hat{\textbf{e}}_i\}$ are classical unit vectors denoting the orientations of the magnetic moments localized at sites $\{i\}$ with magnitudes $\{\mu_i\}$. For positive values of $K_i$ the magnetic moments tend to align along  the easy axis $\hat{\textbf{u}}$.

We remark that Eq.\ (\ref{eq_ham_real}) does not include dipolar interactions nor higher orders of exchange interactions and of anisotropy coefficients. Their inclusion is not required to obtain the extension of the Néel diagram and the corresponding prediction of transitions among homogeneous ferrimagnetic phases in the composition-temperature diagram of GdCo$_5$-type materials presented in this work, since isotropic exchange interactions dominate energetically~\cite{chris,RADWANSKI198657}.

We focus on ferrimagnetic materials with two non-equivalent sites $a$ and $b$ in the unit cell, with  $N_a$ and $N_b$ atomic positions and local magnetic moments $\mu_a$ and $\mu_b$, respectively. For the ferrimagnetic state the magnetic unit cell contains $N_a+N_b$ atoms. The energy per unit cell at zero temperature is
\begin{equation}
\begin{split}
 E_{0K} = & 
    - \gamma_{aa} N_a \hat{\textbf{e}}_a \cdot \hat{\textbf{e}}_a
    - \gamma_{bb} N_b \hat{\textbf{e}}_b \cdot \hat{\textbf{e}}_b
     - \gamma_{ab} N_aN_b \hat{\textbf{e}}_a \cdot \hat{\textbf{e}}_b \\
   &   +K_a N_a \sin^2\theta_a + K_b N_b\sin^2\theta_b \\
  &    -\textbf{B}\cdot (N_a\mu_a \hat{\textbf{e}}_a + N_b\mu_b\hat{\textbf{e}}_b)
    ,
\label{eq_E0K}
\end{split}
\end{equation}
where $K_a$ and $K_b$ are the uniaxial constants, while $\theta_a$ and $\theta_b$ are the angles between moments and easy axes of both sublattices.
$\gamma_{aa}$, $\gamma_{bb}$, and $\gamma_{ab}$ come from the isotropic constants $J_{ij}$, quantifying the exchange interaction for a ferrimagnetic state among and between the two non-equivalent positions, respectively.
Note that at $T=0$~K, $\hat{\textbf{e}}_a \cdot \hat{\textbf{e}}_a=\hat{\textbf{e}}_b \cdot \hat{\textbf{e}}_b=1$ and so the first two terms in the right-hand-side of Eq.\ (\ref{eq_E0K}) are mere constants. We, however, write them down here as the presence of $\gamma_{aa}$ and $\gamma_{bb}$ at finite temperature is important due to local moment fluctuations, see section \ref{methods}.
Moreover, we have found that the magnetic properties of the $2c$ and $3g$ are very similar, which is supported by Monte Carlo simulations as shown in section \ref{MCresults}. This allows the material to be modeled with $N_a=N_\text{TM}=5$ (i.e., $\hat{\textbf{e}}_a=\hat{\textbf{e}}_\text{TM}$) and $N_b=N_\text{Gd}=1$ (i.e., $\hat{\textbf{e}}_b=\hat{\textbf{e}}_\text{Gd}$). These values are taken throughout this work.

\subsubsection{On the origin of the magnetic interactions $\gamma_{aa}$, $\gamma_{bb}$, and $\gamma_{ab}$, and their connection to density functional theory}

The relation between $\{J_{ij}\}$ and $\gamma_\text{TM-TM}$, $\gamma_\text{Gd-Gd}$, $\gamma_\text{TM-Gd}$ is provided by firstly Fourier transforming the interactions
\begin{equation}
J_{ss'} (\textbf{q}) =
\frac{1}{N_{c}} \sum_{tt'} \sum_{\alpha\alpha'} J^{\alpha\alpha'}_{tst's'}
 p_{\alpha} p_{\alpha'}
  e^{-i
 \textbf{q}(\textbf{X}_{ts}-\textbf{X}_{t's'})}
 ,
\label{eq_FT}
\end{equation}
where $N_{c}$ is the total number of unit cells in the crystal, $\textbf{X}_{i}=\textbf{X}_{ts}$ (and $\textbf{X}_{j}=\textbf{X}_{t's'}$) is a vector giving the position of the atoms, and $\textbf{q}$ is the wave vector. Here the lattice indices $i$ and $j$ are split into two indices, $ts$ and $t's'$, respectively. This allows to decompose the position vector as $\textbf{X}_{ts}=\textbf{R}_t+\textbf{r}_s$, where $\textbf{R}_t$ and $\textbf{r}_s$ are vectors pointing to the origin of the unit cell and to the magnetic atom inside this unit cell with respect to its origin, respectively. Therefore, $s$ and $s'$ run over sub-lattice indices while $t$ and $t'$ over the unit cell origins.
$p_\alpha$ is the probability of finding atom $\alpha$ sitting in sublattice $s$. For example, for one of our case studies GdCo$_{5-x}$Ni$_x$, $p_\text{Gd}=1$, $p_\text{Co}=(5-x)/5$ and $p_\text{Ni}=x/5$.
Appendix \ref{App1} shows how to derive the values of $\gamma_\text{TM-TM}$, $\gamma_\text{Gd-Gd}$, and $\gamma_\text{TM-Gd}$ from the elements of matrix $J_{ss'}$ at $\textbf{q}=\textbf{0}$.

The model parameters $\gamma_\text{TM-TM}$, $\gamma_\text{Gd-Gd}$, $\gamma_\text{TM-Gd}$, $\mu_\text{TM}$, and $\mu_\text{Gd}$, together with the uniaxial anisotropy constants, can be either introduced phenomenologically or derived from density-functional theory (DFT) calculations. Two routes can be employed for the latter:
\begin{enumerate}[leftmargin=11pt]
    \item Evaluation of ${J_{ij}}$ via the Liechtenstein method~\cite{lichtenstein}, based on infinitesimal rotations of local moments around a chosen reference magnetic state and application of Eq.\ (\ref{eq_FT}) and derivations given in appendix.
    \item Application of the DFT-based disordered local moment (DLM) framework~\cite{Gyorffy_1985}, which directly yields $\gamma_\text{TM-TM}$, $\gamma_\text{Gd-Gd}$, $\gamma_\text{TM-Gd}$ as Gibbs free energy coefficients in the spirit of a Landau-type expansion~\cite{chris,PhysRevB.105.064425}.
\end{enumerate}

It is worth noting that previous DLM-DFT studies for GdCo$_5$-based ferrimagnets have shown that the terms included in Eq.~(\ref{eq_E0K}) suffice to describe their magnetic behavior~\cite{chris}. Further discussion of their connection to the microscopic Hamiltonian [Eq.~(\ref{eq_ham_real})], as well as methods for constructing and minimizing $\mathcal{G}$, can be found elsewhere~\cite{Edu_Tc,PhysRevB.105.064425,PhysRevLett.118.197202,chris}.

\subsection{Methods}
\label{methods}

In Sect.\ \ref{Neel} we analyze the influence of the parameters $\gamma_\text{TM-TM}$, $\gamma_\text{Gd-Gd}$, $\gamma_\text{TM-Gd}$, $\mu_\text{TM}$, and $\mu_\text{Gd}$ on the Néel diagram using a phenomenological approach. To this end, we construct the general form of the Gibbs free energy $\mathcal{G}$ inspired by the DLM-DFT framework, presented in this section, without performing explicit DLM-DFT calculations. By exploring phenomenological values of the model parameters, we demonstrate how minimizing $\mathcal{G}$ reproduces and extends Néel’s seminal diagram for ferrimagnets. 
We apply these insights to GdCo$_{5}$-type materials, including their ternary and quaternary alloys. To this end, we compute the exchange constants ${J_{ij}}$ using the Liechtenstein method~\cite{lichtenstein} and estimate the associated Gibbs free energy parameters through Eq.~(\ref{eq_FT}), as described in \ref{App1}. The details of these DFT calculations and the subsequent Monte Carlo simulations are given in Secs.~\ref{Jijs} and~\ref{MC}, respectively.

\subsubsection{Mean-field approach and the DLM-based Gibbs free energy}
\label{MF}
At finite temperatures, a mean-field approximation for the Hamiltonian in Eq.\ (\ref{eq_ham_real}) can be implemented via the Peierls-Feynmann inequality~\cite{PhysRev.97.660,doi:10.1143/JPSJ.50.1854}. This requires to introduce a trial Hamiltonian, which is chosen to describe a coupling of the localized local moment orientations $\{\hat{\textbf{e}}_\text{TM}, \hat{\textbf{e}}_\text{Gd}\}$ with site-dependent molecular fields $\{\textbf{h}_\text{TM}, \textbf{h}_\text{Gd}\}$~\cite{Gyorffy_1985}:
\begin{equation}
    \mathcal{H}_\text{tr} = -N_\text{TM}\textbf{h}_\text{TM} \cdot \hat{\textbf{e}}_\text{TM}-\textbf{h}_\text{Gd} \cdot \hat{\textbf{e}}_\text{Gd}
    .
\label{eq_Htr}
\end{equation}
The probability of different local moment configurations associated with $\mathcal{H}_\text{tr}$ factorizes due to the single-site nature of the trial Hamiltonian to 
\begin{equation} P(\{\hat{\textbf{e}}_\text{TM}, \hat{\textbf{e}}_\text{Gd}\})=\prod_t\left[P_{t,\text{TM}}(\hat{\textbf{e}}_\text{TM})\right]^{N_\text{TM}}P_{t,\text{Gd}}(\hat{\textbf{e}}_\text{Gd}) \;.\end{equation} 
Here the index $t$ labels the unit cells throughout the crystal and  the single-site probabilities are given as
\begin{equation}
    P_i(\hat{\textbf{e}}_i)=\frac{1}{\mathcal{Z}_{\text{tr},i}}\exp\left[\beta\textbf{h}_i\cdot\hat{\textbf{e}}_i\right]
    ,
\label{eq_Pi}
\end{equation}
where the normalization $\mathcal{Z}_{\text{tr},i}=\frac{4\pi\sinh(\beta h_i)}{\beta h_i}$  comes from the calculation of the partition function, and $\beta=\frac{1}{k_\text{B} T}$ is the Boltzmann factor.

The Peierls-Feynmann inequality provides a mean-field based Gibbs free energy that minimizes with respect to the molecular field parameters $\textbf{h}_\text{TM}$ and $\textbf{h}_\text{Gd}$, $\mathcal{G}= -TS+H$,~\cite{Edu_Tc,PhysRevB.105.064425}
where $S=N_\text{TM}S_\text{TM}+S_\text{Gd}$ is the total entropy composed of sublattice entropies
\begin{equation}
\begin{split}
  &  S_i  =
    k_\text{B}\langle-\log P_i\rangle_\text{tr} \\
 &  =k_\text{B}\left[
    1+\ln\left(4\pi\frac{\sinh(\beta h_i)}{\beta h_i}\right)
    -\beta h_i\coth(\beta h_i)
    \right]
,
\label{eq_S}
\end{split}
\end{equation}
$\langle\dots\rangle_\text{tr}$ denoting the thermal average using the trial probability in Eq.\ (\ref{eq_Pi}). $H$ is the enthalpy, that is, the generalization of Eq.\ (\ref{eq_E0K}) to finite temperature obtained by performing the thermal average of Eq.\ (\ref{eq_ham_real}).
The corresponding mean-field Gibbs free energy, basis of DLM-DFT theory for magnetic phase transitions, thus is
\begin{equation}
\begin{split}
    \mathcal{G}
    = &
    -T[5S_\text{TM}(\beta \textbf{h}_\text{TM})+S_\text{Gd}(\beta \textbf{h}_\text{Gd})]
    +5K_\text{TM}f_u(\beta \textbf{h}_\text{TM}) \\
    &
    -5\gamma_\text{TM-Gd} \textbf{m}_\text{TM}\cdot\textbf{m}_\text{Gd}  -5\gamma_\text{TM-TM} m_\text{TM}^2
    \\
    &
    -\gamma_\text{Gd-Gd} m_\text{Gd}^2
    -\textbf{B}\cdot\left(5\mu_\text{TM}\textbf{m}_\text{TM} + \mu_\text{Gd}\textbf{m}_\text{Gd}\right)
    ,
\label{eq_G}
\end{split}
\end{equation}
where
\begin{equation}
    \textbf{m}_i = \langle \hat{\textbf{e}}_i \rangle_\text{tr}=\left[-\frac{1}{\beta h_i} + \coth(\beta h_i) \right] \hat{\textbf{h}}_i
\label{eq_m}
\end{equation}
are local magnetic order parameters, while the thermal averages of the anisotropy terms are
\begin{equation}
\begin{split}
 &  f_u(\beta \textbf{h}_i) = \langle \sin^2(\theta_i) \rangle_\text{tr}
  = 1 - \left(\hat{\textbf{h}}_i\cdot \hat{\textbf{z}}\right)^2  \\
   & 
    +\frac{1}{\beta h_i}\left(\frac{-1}{\beta h_i}+\coth(\beta h_i)\right)\left(3(\hat{\textbf{h}}_i\cdot \hat{\textbf{z}})^2-1\right)
    .
\label{eq_m}
\end{split}
\end{equation}

The molecular fields $\textbf{h}_\text{TM}$ and $\textbf{h}_\text{Gd}$ sustain magnetic ordering at finite temperature. Their values provide the dependence on temperature of the magnetic entropies, $S_\text{TM}$ and $S_\text{Gd}$, the order parameters, 
$\textbf{m}_\text{TM}(\beta \textbf{h}_\text{TM})$ and $\textbf{m}_\text{Gd}(\beta \textbf{h}_\text{Gd})$, and the uniaxial anisotropy function, $f_u$, such that all of these quantities depend on $\beta\textbf{h}_\text{TM}$ and $\beta\textbf{h}_\text{Gd}$ following the mean-field approximation. $\mu_\text{TM}$ and $\mu_\text{Gd}$ are the corresponding averaged local moment magnitudes, $\textbf{M}_\text{tot}=\mu_\text{TM}\textbf{m}_\text{TM}+\mu_\text{Gd}\textbf{m}_\text{Gd}$ being the total magnetization.
The temperature-dependent magnetic properties  are obtained by minimizing Eq.\ (\ref{eq_G}) with respect to $\{\textbf{m}_i\}$ (or equivalently $\{\textbf{h}_i\}$) at different values of $T$ and $\textbf{B}$ with respect to the order parameters and self-consistently with $\textbf{h}_\text{TM}$ and $\textbf{h}_\text{Gd}$~\cite{chris,julie,Edu_Tc,PhysRevB.105.064425,PhysRevLett.118.197202}.

The DLM-DFT picture assumes a separation of time scales between fast electronic motion and slower transverse fluctuations of robust local moments. Thermodynamic averages are, in consequence, performed over local-moment orientation configurations. It is therefore appropriate for systems with well-defined magnetic moments, such as magnetic materials containing rare-earth elements: GdCo$_5$-type compounds~\cite{chris}, the heavy rare-earth elements~\cite{PhysRevLett.118.197202}, and Eu$_2$In~\cite{PhysRevB.101.174437}. The DLM-DFT framework has also been successfully applied to other non-rare-earth materials, including FeRh~\cite{julie}, FeAl~\cite{halil_prm}, B20-MnGe~\cite{PhysRevB.103.024410}, as well as Mn$_3$A~\cite{PhysRevB.105.064425} and Mn$_3$AN (A = Ga, Ni, Sn) triangular antiferromagnets~\cite{PhysRevX.8.041035,PhysRevB.95.184438}.

\subsubsection{Calculation of the coercive field}

A ferrimagnetic ground state 
is stabilized by $\gamma_\text{TM-Gd}<0$, $\gamma_\text{TM-TM}>0$, $\gamma_\text{Gd-Gd}>0$. Its net magnetization is given by $\textbf{M}_\text{tot}=5\mu_\text{TM}\textbf{m}_\text{TM}+\mu_\text{Gd}\textbf{m}_\text{Gd}$.
The coercive field is defined as the magnetic field $\textbf{B}_c$, applied antiparallel to $\textbf{M}_\text{tot}$, at which the state ceases to be a local minimum~\cite{Ostler}.
At finite temperature, $\textbf{B}_c$ must be calculated numerically. However, in the zero-temperature limit an analytical expression can be obtained by analyzing the Hessian matrix of $E_{0K}$ with respect to the two angular variables:
\begin{equation}
    \textbf{B}_{c}(T=0\text{ K}) = 
    \frac{-b-\sqrt{b^2 -4\mu_\text{TM}\mu_\text{Gd} c}}{2\mu_\text{TM}\mu_\text{Gd}}
    (-\hat{\textbf{M}})
\label{eq_Bc0K}
\end{equation}
where $b=\gamma_\text{TM-Gd}(N_\text{TM}\mu_\text{TM}-\mu_\text{Gd})-2(K_\text{Gd}\mu_\text{TM}-K_\text{TM}\mu_\text{Gd})$ and $c=-4K_\text{TM}K_\text{Gd}+2\gamma_\text{TM-Gd}(N_\text{TM}K_\text{TM}+K_\text{Gd})$ if $N_\text{TM}\mu_\text{TM}>\mu_\text{Gd}$. 
Importantly, the coercive field is independent of the magnetic intralattice exchange ($\gamma_\text{TM-TM}$ and $\gamma_\text{Gd-Gd}$).
If instead $N_\text{TM}\mu_\text{TM}<\mu_\text{Gd}$, the coefficient $b$ changes sign. This sign reversal leads to a discontinuity in the coercive field when $\textbf{M}_\text{tot}$ flips direction.

We emphasize that dipolar interactions, microstructure, and thermally activated domain-wall and nucleation processes can influence magnetization reversal mechanisms and experimentally measured coercive fields. Consequently, the switching fields obtained within the present homogeneous free-energy framework should be interpreted as intrinsic instability fields of the ferrimagnetic state rather than quantitative predictions of experimental coercivity.

\subsubsection{\textit{Ab initio} calculation of magnetic interactions}
\label{Jijs}

We calculate  the isotropic pairwise interactions $\{J_{ij}\}$ using the DFT-based Liechtenstein method~\cite{lichtenstein}, as implemented in the AkaiKKR package~\cite{Akai_4,Akai_1}, which employs the Korringa–Kohn–Rostoker (KKR) electronic structure method~\cite{krringa,kohn_rostoker}. The atomic sphere approximation (ASA) with values of the maximum angular momenta of $l_\text{max} = 3$ and $l_\text{max} = 2$ for the single-site scattering problem have been used for the Gd and TM sites, respectively. 
Our interest are the isotropic pairwise interactions so we stick to scalar relativistic calculations.
We use the local density approximation as implemented by Moruzzi, Janak, and Williams \cite{moruzzi} in combination with the ``open core" scheme, in which the electronic structure is constrained to 
half-filled 4\textit{f}-shell to fulfill Hund's rules.

In addition to the pristine material GdCo$_{5}$, alloys of GdCo$_{5-x}$TM$_x$ with TM:  Ni, Cu, and Fe, as well as quaternary GdCo$_{5-x}$(Fe$_{0.5x}$Ni$_{0.5x}$), have been studied by means of the coherent potential approximation (CPA)~\cite{cpa_main_ref} assuming a homogenous distribution of the substituents on $2c$ and $3g$ crystallographic sites.
Previous calculations have shown that the magnetic moment magnitudes are only weakly dependent on the location of TM dopants in the case of Ni, while their influence on the exchange constants may be more substantial~\cite{amy}, an effect that we leave for future investigations.
The experimental lattice parameters of GdCo$_5$ from literature \cite{chuang,lat_prm_gdco5} have been used for all material compositions.

\begin{figure}[t]
    \centering
    \includegraphics[width=0.45\textwidth]{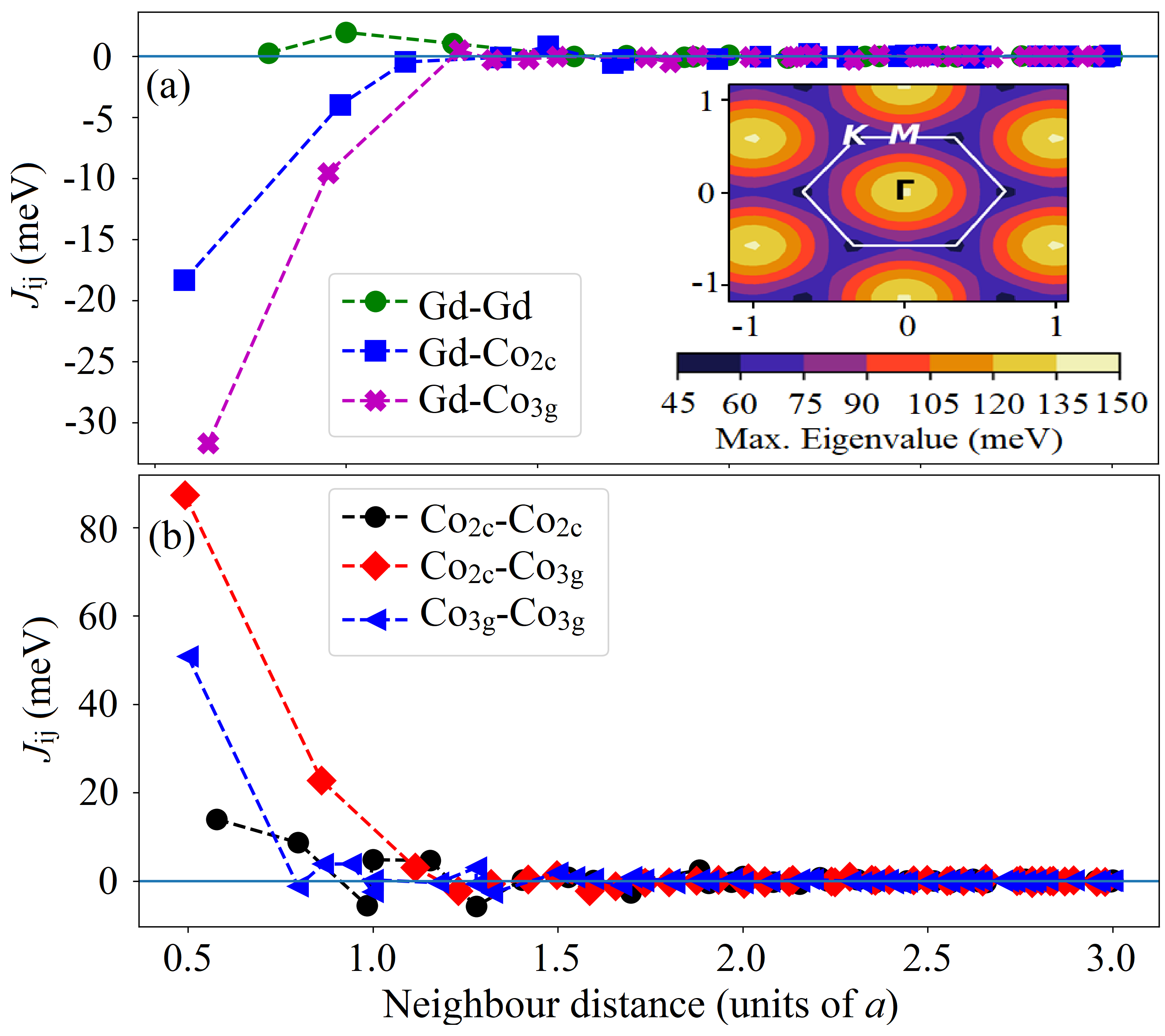}
    \caption[Exchange interaction parameters in GdCo$_{5}$.]{(a,b) Magnetic interactions \textit{$J^{\alpha\alpha'}_{ij}$} computed for GdCo$_{5}$ as functions of the distance between the two atoms (in units of lattice constant $a$). The Gd-Gd and Gd-Co interactions are shown in panel (a), while the Co-Co  interactions are shown in panel (b).
    The inset shows the largest eigenvalue of the Fourier transform of the interaction matrix, $J^{\alpha\alpha'}_{ss'}(\textbf{q})$, against the wave vector, see Eq.\ (\ref{eq_FT}). Characteristic reciprocal lattice points of the Brillouin zone of a hexagonal close packed structure are shown (K, M, and $\Gamma$).
    }
    \label{fig:jij_GdCo5}
\end{figure}

Because the isotropic real-space parameters depend on the atomic species occupying the TM sites, they carry an additional index $\alpha$ ($\alpha'$) that specifies the substitutional element at site $i$ ($j$). Thus, the exchange interactions are written as ${J^{\alpha\alpha'}_{ij}(x)}$, where, for example, $\alpha =$ Co, Ni, Cu, or Fe at the ${2c}$ and ${3g}$ sites.
The distance dependency of the interactions are exemplified for GdCo$_5$ in Fig.~\ref{fig:jij_GdCo5}. Importantly, these decay fast with the distance, being negligible up to the chosen cutoff of three lattice constants.
The ferromagnetic Co-Co  are the dominant interactions, $\gamma_\text{Gd-Gd}$ therefore being much smaller than $\gamma_\text{TM-TM}$ and $\gamma_\text{Gd-TM}$ in Eq.\ (\ref{eq_G}) [see Eq.\ (\ref{eq_JtoU})].

The Fourier transform $J^{\alpha\alpha'}_{ss'} (\textbf{q})$, see Eq.\ (\ref{eq_FT}), allows to  analyze the spectrum of favorable magnetic states. The inset of Fig.\ \ref{fig:jij_GdCo5} shows the largest eigenvalue of $J_{ss'}(\textbf{q})$, which gives the dominant instability of the paramagnetic state~\cite{Edu_Tc,PhysRevB.103.024410}.
This value peaks at $\textbf{q}=\textbf{0}$ with eigenvector components describing antiparallel alignment between Gd and TM atoms and confirms the stability of the ferrimagnetic state.

\subsubsection{Monte Carlo simulations}
\label{MC}

Monte Carlo simulations for the Heisenberg Hamiltonian described in Eq.\ (\ref{eq_ham_real}) containing the DFT-based $\{J_{ij}^{\alpha\alpha'}\}$ are performed to validate our mean-field predictions. We employ the Mote Carlo code developed by Mavropoulos and coworkers \cite{phivos_mc_1,phivos_mc_2}, designed to account for the effect of chemical disorder at each lattice site in alignment with the CPA.
Anisotropy effects are not included within these Monte Carlo simulations.
We construct $12 \times 12 \times 12$ supercells, i.e.\ 19384 atoms. For each temperature we first equilibrate the system during 1000 Monte Carlo steps and then sample the thermodynamic properties for 5000 steps taking every 25 step into account.
The phase transition temperature is determined by monitoring the peak of the heat capacity
$C_{V}^\text{mag} = \frac{1}{k_B T^2}
    \left(
    \langle E^2 \rangle - \langle E \rangle^2
    \right)$,
where $E$ is the internal magnetic energy and $\langle\dots\rangle$ denotes the average over Monte Carlo simulations.

\subsubsection{Electronic band structure and Fermi surface calculation}

The electronic band-structure and Fermi Surface calculations presented in Section~\ref{results} were also performed using the KKR method, as implemented in the Hutsepot code~\cite{Ebert_2011,Däne_2009}, combined with the CPA to account for chemical disorder, consistent with the treatment employed in the calculations of the magnetic interactions.

\section{Results}
\label{results}

\subsection{Extension of the Néel diagram}
\label{Neel}

\begin{figure}[t]
    \centering
    \includegraphics[scale=.72]{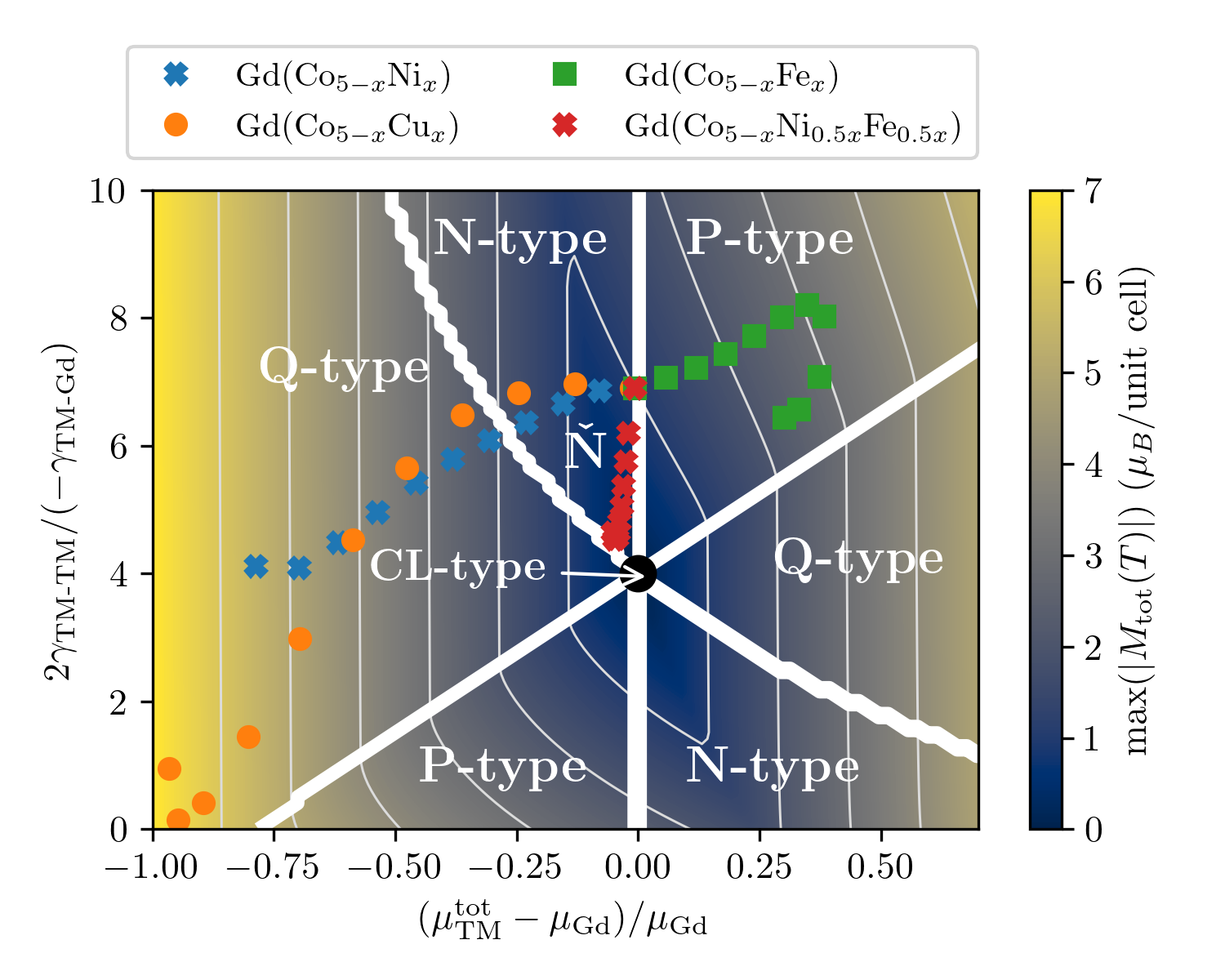}
    \caption{
    Extended Néel diagram for the total net magnetic moment at zero temperature and the ratio of inter and intralattice interaction $\gamma_\text{TM-TM}/\gamma_\text{TM-Gd}$.
All phases meet at the fully compensated line.
   The color bar indicates the maximum value of the total absolute magnetization.
DFT data for ternary and quaternary GdCo$_5$-type ferrimagnets are also shown. All data sets originate from $x=0$, corresponding to pristine GdCo$_5$. Data points departing from this common origin correspond to progressively larger values of $x$ in increments of $\Delta x=0.5$, see legend.}
    \label{fig:diagram1}
\end{figure}

We first conduct a general exploration of the ferrimagnetic behaviors that emerge from minimizing $\mathcal{G}$ over representative parameter choices, including but not limited to those derived from DFT.

The Néel diagrams in Ref.~\cite{Smart} are restricted to cases where the total, zero-temperature, magnetic moment of one sublattice is substantially larger than that of the other, in particular $\mu_a^\text{tot}=2\mu_b^\text{tot}$. Q, P, and N curves shown in Fig.~\ref{fig:intro}(a) correspond to this typical ferrimagnetic behavior. Consequently, the axes of the diagrams in Ref.~\cite{Smart} were chosen to represent the remaining relevant parameters given by the ratios of effective exchange interactions, i.e., $\gamma_\text{TM-TM}/\gamma_\text{TM-Gd}$ and $\gamma_\text{Gd-Gd}/\gamma_\text{TM-Gd}$.
However, this representation is insufficient to explore the potential stabilization of materials with $\mu_\text{TM}^\text{tot}\approx\mu_\text{Gd}$, an unexplored and physically interesting case that we address in this work.
Upon continuously tuning the material parameters, an extreme limit of the N-type behavior emerges, characterized by an exceptionally small net magnetization over a broad temperature interval around the compensation point.
Further tuning may stabilize a ferrimagnetic phase consisting of a continuous line of compensation points, which we refer to as the CL-type phase, see Fig.~\ref{fig:intro}(a).

Our extension of Néel’s diagram thus follows by analyzing the different types of ferrimagnetic curves as a function of $\gamma_\text{TM-TM}/\gamma_\text{TM-Gd}$ ($y$-axis) and, crucially, the total net magnetic moment at zero temperature ($x$-axis). The resulting diagram is presented in Fig.~\ref{fig:diagram1}.
The effect of $\gamma_\text{Gd-Gd}$ has been neglected. This applies to GdCo$_5$-type ferrimagnets, where the interactions satisfy $\gamma_\text{Gd-Gd}\ll \gamma_\text{TM-TM}$ and $\gamma_\text{Gd-Gd}\ll 5|\gamma_\text{TM-Gd}|$. For example, in GdCo$_5$ the respective DFT values that we have computed are $\gamma_\text{Gd-Gd}=6.3$ meV, $\gamma_\text{TM-TM}=264$ meV, and $5\gamma_\text{TM-Gd}=-190$ meV.
Fig.~\ref{fig:diagram1} also displays the DFT data for the GdCo$_5$-type ferrimagnets under consideration. For $x = 0$, all datasets corresponding to the three material series studied, Gd(Co$_{5-x}$TM$_x$) with TM = Ni, Fe, or Cu, as well as Gd(Co$_{5-x}$Ni$_{0.5x}$Fe$_{0.5x}$), originate from the same reference point, i.e., the pristine GdCo$_5$. Data points progressively departing from this initial point correspond to increasing values of $x$.

These effective free-energy parameters arise from the connection between DLM-inspired free-energy expansions and exchange-interaction matrices, as also done in Ref.~\cite{PhysRevB.103.024410}. The close agreement with atomistic Monte Carlo simulations presented in Sec.\ \ref{MCresults} indicates that the resulting phase boundaries are robust for ferrimagnets describable by two competing antiparallel effective sublattices.

It is noteworthy that transitions between any two ferrimagnetic types are possible. For instance, a P-type phase may transform into either a Q-type or an N-type phase through the emergence of a low-temperature maximum in $M_\text{tot}(T)$ or by the appearance of a compensation point at $T_\text{tr}$, respectively.
When the total magnetic moment on the TM sites exceeds that on the rare-earth site ($\mu^\text{tot}_\text{TM} > \mu_\text{Gd}$), larger values of $\gamma_\text{TM–TM}$ relative to $|\gamma_\text{TM–Gd}|$ favor the stabilization of a P-type phase, while an N-type phase is stabilized when $\mu^\text{tot}_\text{TM} < \mu_\text{Gd}$. Opposite trends are observed for large negative values of $\gamma_\text{TM–TM}$. Moreover, the Q-type phase remains stable for intermediate values of $\gamma_\text{TM–TM}$.

The CL-type phase corresponds to a diagram point at which all phase boundaries converge, which is generally given by (see appendix \ref{App1})
\begin{equation}
    \frac{2(\gamma_\text{TM-TM}-\gamma_\text{Gd-Gd})}{-\gamma_\text{TM-Gd}}=N_\text{TM}-N_\text{Gd}
    \label{eq_cond}
\end{equation}
under the condition $\mu^\text{tot}_\text{TM} = \mu_\text{Gd}$.
Recall that $\gamma_\text{Gd-Gd}\approx0$, $N_\text{TM} = 5$ and $N_\text{Gd} = 1$ for GdCo$_5$-type ferrimagnets, so this diagram point occurs at $2\gamma_\text{TM–TM}/|\gamma_\text{TM–Gd}| = 4$ in this materials class.
Approaching the CL-type condition brings the system closer to a fully compensated phase at all temperatures [dashed black line in Fig.~\ref{fig:intro}(a)].
The N-type ferrimagnetic configuration that stabilizes in the vicinity of this diagram point has a very small net magnetization over a broad temperature range around its single compensation point.

Fig.~\ref{fig:diagram1} shows that the difference $\mu^\text{tot}_\text{TM} - \mu_\text{Gd}$ is negative and decreases further with increasing $x$ in Gd(Co$_{5-x}$Ni$_x$) and Gd(Co$_{5-x}$Cu$_x$). This is because Ni and Cu substitution reduces the total magnetic moment at the TM sites. In contrast, substitution with Fe, which carries a robust local moment, markedly enhances $\mu^\text{tot}_\text{TM}$. This observation motivates the exploration of double co-substitution with Ni and Fe, Gd(Co$_{5-x}$Ni$_{0.5x}$Fe$_{0.5x}$), enabling access to the central region of the extended Néel diagram containing the fully compensated line. Indeed, a N-type phase close to the CL diagram point can be stabilized in this quaternary alloy because, for GdCo$_5$ ($x = 0$), $\mu^\text{tot}_\text{TM} \lessapprox \mu_\text{Gd}$, and increasing $x$ preserves this balance: the reduction of the TM moment due to Ni is compensated by the enhancement from Fe. At the same time, the intra-sublattice interaction $\gamma_\text{TM–TM}$ weakens relative to $\gamma_\text{TM–Gd}$ (although both decrease in magnitude), thereby favoring the stabilization of the N-type state closer to the fully compensated line.

We emphasize that the CL-type condition in Eq.\ (\ref{eq_cond}) is generally applicable to ferrimagnets modeled by two magnetic sublattices. Gd(Co$_{5-x}$Ni$_{0.5x}$Fe$_{0.5x}$) remains unsynthesized but, notably, Stinshoff \textit{et al}.\ have experimentally realized a slightly overcompensated sample with a composition very close to that of the quaternary Mn$_{1.5}$FeV$_{0.5}$Al Heusler ferrimagnet. This material exhibits an N-type configuration near the fully compensated line, with a total magnetization below 0.06 $\mu_B$/f.u.\ at all temperatures. This behavior was achieved through Fe and V substitution at the Mn $8c$ and $4b$ sites~\cite{Stinschoff_PhysRevB.95.060410}. Recent theoretical studies of L-type ferrimagnets, which are fully compensated at 0 K, have advanced this line of research further by predicting multiple compensation points in Mn$_{2.1}$Co$_{0.45}$V$_{0.45}$Al, which is likewise expected to lie near the fully compensated line~\cite{l913-4x5y}.

\begin{figure}[t]
    \centering
    \includegraphics[scale=.70]{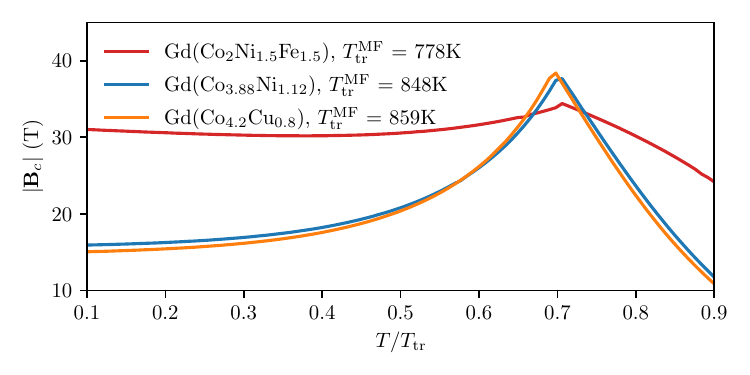}
    \caption{Temperature dependence of coercive fields calculated for selected GdCo$_5$-type ferrimagnets.
    The temperature axis is normalized to the transition temperature to the paramagnetic state, $T_\text{tr}$.}
    \label{fig:Bcall}
\end{figure}

\subsection{The coercive field}
Although the N-type ferrimagnetic configuration close to the fully compensated line exhibits lower net magnetization values at all temperatures, the behavior of its coercive field is not obvious. To address this, we calculate the temperature dependence of the intrinsic switching field by minimizing the mean-field Gibbs free energy given in Eq.~(\ref{eq_G}) under the effect of an applied field. Fig.~\ref{fig:Bcall} presents the results for selected compositions chosen such that the ratio between the compensation temperature and the transition temperature to the paramagnetic state remains approximately constant across all materials, e.g., $T_\text{comp}/T_\text{tr} \approx 0.7$.
The selected systems are Gd(Co$_2$Ni$_{1.25}$Fe$_{1.25}$) ($x=3$), which shows a very small magnetization similar to the \v{N}-type curve in Fig.~\ref{fig:intro}(b), as well as Gd(Co$_{3.88}$Ni$_{1.12}$) ($x=1.12$) and Gd(Co$_{4.2}$Cu$_{0.8}$) ($x=0.8$), both representative N-type ferrimagnets. The mean-field values of $T_\text{tr}$ are similar for all three compositions, providing a suitable basis for comparison. 
The anisotropy constants have been set to $K_\text{Gd}=0$ and $K_\text{TM}=0.8$ meV following~\cite{RADWANSKI19921321,chris,RADWANSKI198657}.

As shown in Fig.\ \ref{fig:Bcall}, the intrinsic switching field peaks when the total net magnetization vanishes at $T_\text{comp}$, which is in agreement with experimental and theoretical findings in rare earth-TM$_5$ compounds~\cite{10.1063/1.322787,chuang,Nora_GdCoCu,DEOLIVEIRA20111890,amy}.
Due to missing temperature and microstructure and the so-called Brown's paradox, e.g.\ missing domain wall motion and thermally activated nucleation, the absolute values for $B_c$ cannot be compared to experiment.

Interestingly, and somewhat unexpectedly, the estimated $B_c$ is slightly smaller at $T_\text{comp}$ for Gd(Co$_2$Ni$_{1.25}$Fe$_{1.25}$) than for the typical N-type phases. However, Gd(Co$_2$Ni$_{1.25}$Fe$_{1.25}$) exhibits enhanced intrinsic switching field not only in the vicinity of $T_\text{comp}$ but also over a broad temperature range extending from low temperatures up to near $T_\text{tr}$. 
$B_c$ remains large and roughly flat below $T_\text{comp}$, as also observed experimentally for Mn$_{1.5}$FeV$_{0.5}$Al~\cite{Stinschoff_PhysRevB.95.060410}.
Therefore, enhanced intrinsic switching fields, and potentially enhanced coercivity, can be achieved in N-type ferrimagnets over much larger temperature spans, effectively tuned through controlled chemical substitution.

\subsection{DFT-informed Monte Carlo simulations and composition-temperature phase diagrams}
\label{MCresults}

\begin{figure}[t]
    \centering
    \includegraphics[scale=.08]{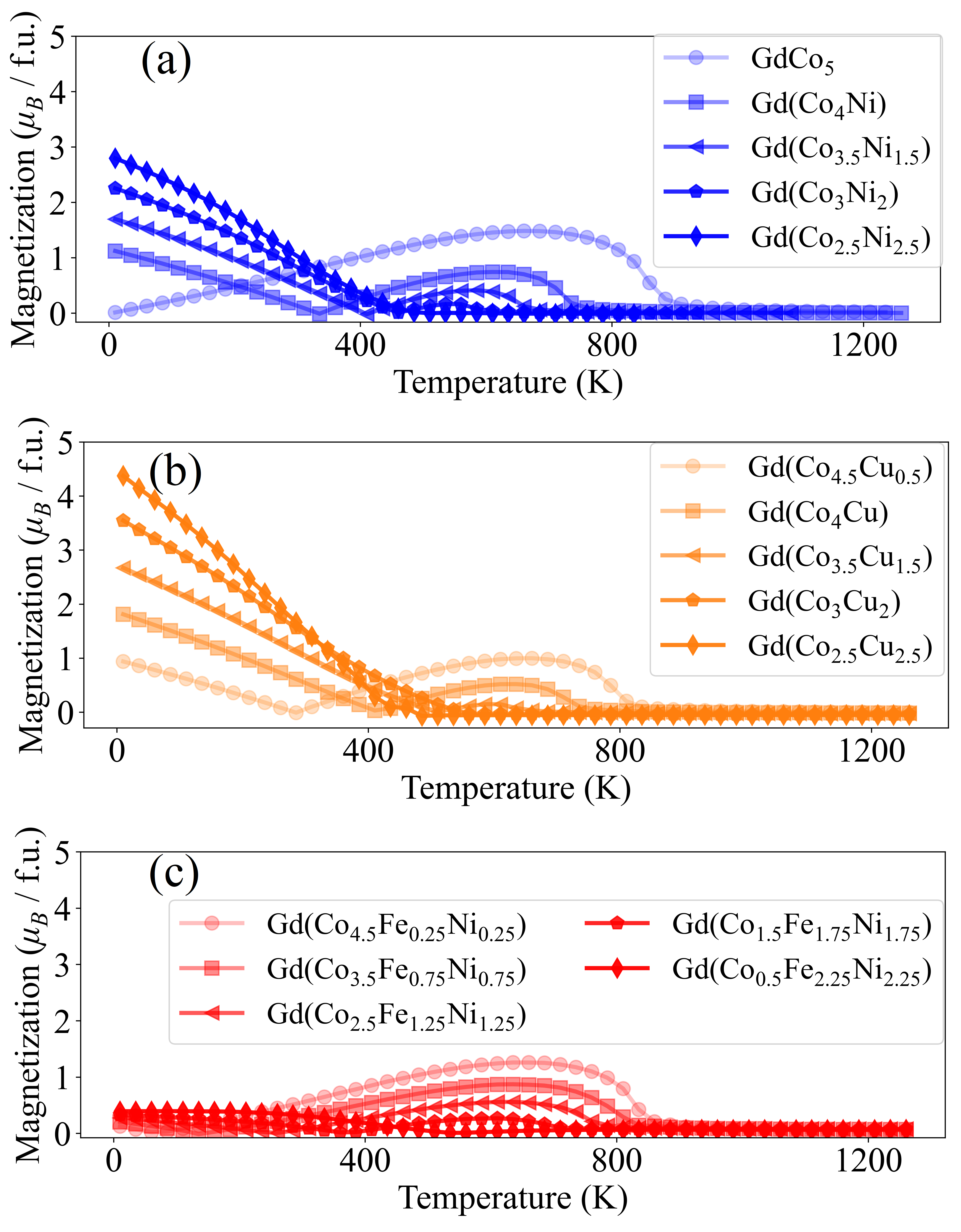}
    \caption{Temperature dependence of total magnetization of (a) Gd(Co$_{5-x}$Ni$_{x}$), (b) Gd(Co$_{5-x}$Cu$_{x}$), and (c) Gd(Co$_{5-x}$Ni$_{0.5x}$Fe$_{0.5x}$) for increasing values of $x$ from Monte Carlo simulations.}
    \label{fig:mag_fcore}
\end{figure}
\begin{figure}[t]
    \centering
    \includegraphics[scale=.07]{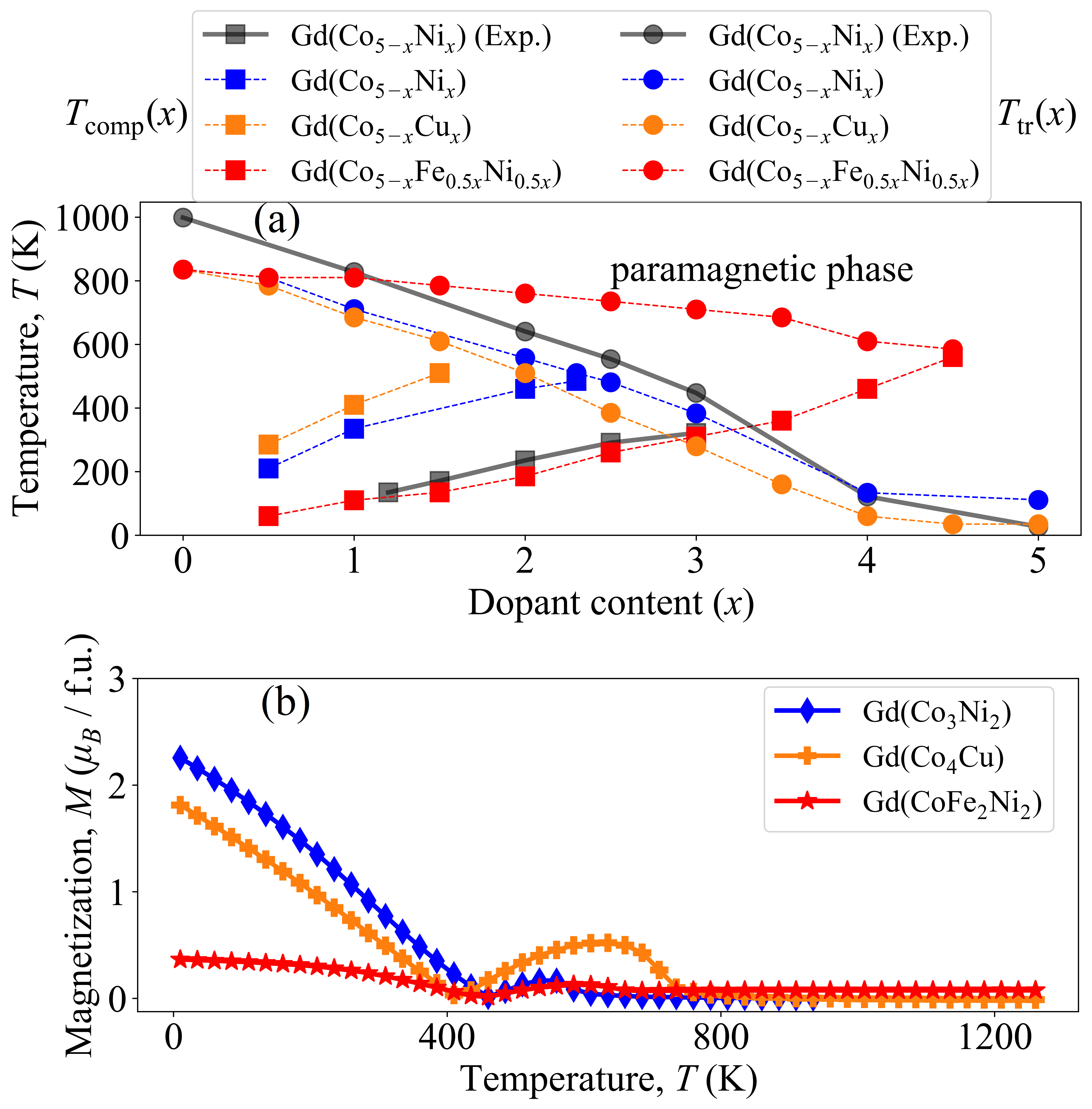}
    \caption{(a) Composition-temperature magnetic phase diagrams obtained for Gd(Co$_{5-x}$Ni$_{x}$), GdCo($_{5-x}$Cu$_{x}$), Gd(Co$_{5-x}$Fe$_{x}$), and Gd(Co$_{5-x}$Fe$_{0.5x}$Ni$_{0.5x}$) by means of Monte Carlo simulations. 
    The compensation temperature, $T_\mathrm{comp}(x)$, and the transition temperature from the paramagnetic state, $T_\text{tr}(x)$, are indicated with squares and circles, respectively.
    Available experimental data for Gd(Co$_{5-x}$Ni$_{x}$)~\cite{amy,chuang} are also shown in black. 
    (b) Temperature dependence of the total net magnetization for selected compositions of interest.
    }
\label{fig:combined_pdf_all}
\end{figure}

The DFT data plotted in Fig.~\ref{fig:diagram1} were obtained by mapping the isotropic exchange parameters $J_{ij}$ onto the mean-field model through Eq.~(\ref{eq_JtoU}). We now show the results of Monte Carlo simulations for $J_{ij}$ to validate our mean-field predictions.

Results are presented in Fig.~\ref{fig:mag_fcore}. Pristine GdCo$_5$ lies close to the boundary separating the N-type and P-type phases, as indicated by its nearly compensated total magnetization $M_\text{tot} \approx 0$ at $T = 0$~K [see the light-colored curve in panel~(a)]. Substitution in Gd(Co$_{5-x}$TM$_x$), with TM = Ni or Cu, initially leads to an increase in the compensation temperature $T_\text{comp}$, accompanied by a reduction in the maximum of $M_\text{tot}(T)$ in the range $T_\text{comp} < T < T_\text{tr}$. At higher substitution levels, this trend culminates in an N-to-Q transition, where the compensation point disappears at $T_\text{tr}$, in agreement with previous experimental and theoretical studies~\cite{amy}.

Fig.~\ref{fig:combined_pdf_all}(a) presents the composition–temperature magnetic phase diagrams constructed from these Monte Carlo simulations, which shows the dependencies of $T_\text{comp}$ and $T_\text{tr}$ on $x$.
Available experimental data for the case of Gd(Co$_{5-x}$Ni$_x$) is also plotted, demonstrating good agreement between theory and experiment at both qualitative and quantitative levels.
The changes observed in the Cu-substituted series, Gd(Co$_{5-x}$Cu$_x$), progress more rapidly than in the Ni-substituted counterpart, reflecting the smaller perturbation introduced by Ni, whose valence configuration is closer to Co's. The critical substitution levels at which the N-to-Q transition occurs are approximately $x \approx 2.5$ for TM = Ni and $x \approx 1.5$ for TM = Cu. In contrast, no such transition is observed for Gd(Co$_{5-x}$Ni$_{0.5x}$Fe$_{0.5x}$).
These results are in good agreement with the mean-field mapping of the DFT data shown in the extended Néel diagram in Fig.~\ref{fig:diagram1}, which supports the reduction of the inequivalent $2c$ and $3g$ sites to a single effective TM sublattice.

Fig.~\ref{fig:combined_pdf_all}(b) shows the total net magnetization for selected values of $x$ across the three alloy series that display N-type behavior. All of them exhibit similar compensation temperatures around $T_\text{comp} \approx 400$~K, analogous to the case investigated in Fig.~\ref{fig:Bcall}. While Gd(Co$_{5-x}$Ni$_x$) and Gd(Co$_{5-x}$Cu$_x$) are typical N-type curves, the quaternary alloy Gd(Co$_{5-x}$Ni$_{0.5x}$Fe$_{0.5x}$) exhibits very small net magnetization over the entire temperature range, in agreement with the mean-field analysis.

\subsection{Non-relativistic exchange spin splitting}
\label{alter}

\begin{figure}[t]
    \centering
    \includegraphics[scale=.68]{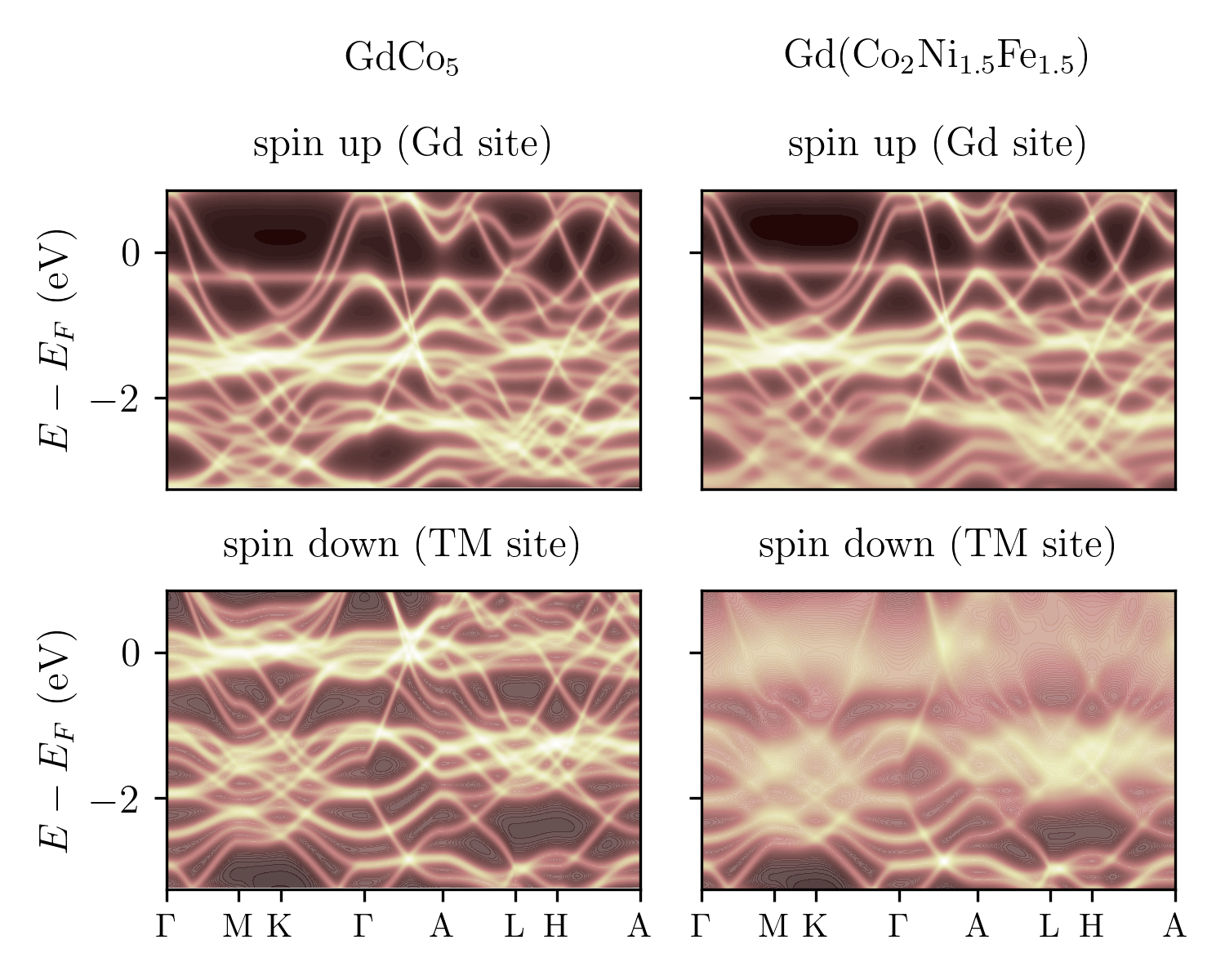}
    \caption{Electronic band structures calculated for GdCo$_5$ (left) and Gd(Co$_2$Ni$_{1.5}$Fe$_{1.5}$) (right) at zero temperature.
    No spin-orbit coupling is included.}
    \label{fig:bands}
\end{figure}

\begin{figure}[t]
    \centering
    \includegraphics[scale=0.85]{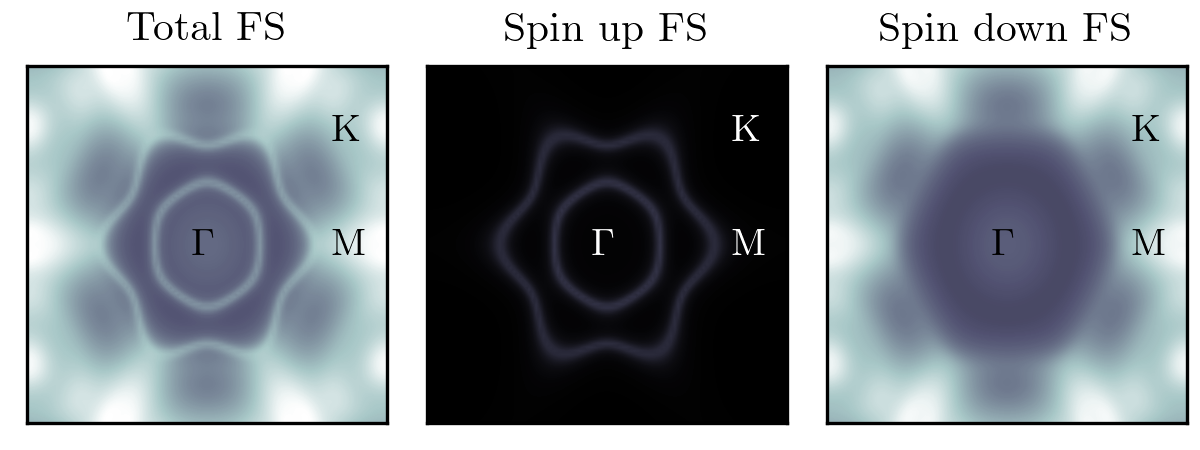}
    \caption{Total Fermi surface (FS) plot and its spin up and down components cut at $k_z=0$. Characteristic reciprocal lattice points of the Brillouin zone of a hexagonal close packed structure are shown (K, M, and $\Gamma$).}
    \label{fig:FS}
\end{figure}

Ferrimagnetic compounds very close or at the diagram point corresponding to the fully compensation line exhibit a vanishing small net magnetization coexisting with a strong spin splitting in the reciprocal space of purely non-relativistic origin. Figure \ref{fig:bands} shows the DFT-computed spin-up (Gd) and spin-down (TM) projected bands for GdCo$_5$-type ferrimagnets. A large spin splitting of order $\sim1$ eV (most clearly visible for GdCo$_5$, left panels) arises because TM's and Gd's electrons belong to different chemical sublattices. We note that Gd(Co$_2$Ni$_{1.5}$Fe$_{1.5}$), which is very close to the fully compensation line, has an almost vanishing total magnetic moment (below $|\mu_\text{tot}|= 0.04\mu_B$ per primitive cell at all temperatures), while its spin splitting remains essentially identical to that of GdCo$_5$. The observed band broadening is caused by the chemical disorder.
For completeness, Fig.\ \ref{fig:FS} displays the corresponding Fermi surface of a representative GdCo$_5$-type compound, providing a complementary view of the electronic structure.

A variety of altermagnetic materials have been extensively studied in recent years, including CrSb~\cite{Yang2025,Reimers2024,PhysRevLett.133.206401,Li2025,https://doi.org/10.1002/advs.202502226}, MnTe~\cite{PhysRevLett.133.156702,Amin2024,PhysRevLett.130.036702}, and RuO$_2$~\cite{doi:10.1126/sciadv.aaz8809,Zhang2025,PhysRevLett.126.127701,PhysRevLett.128.197202,PhysRevLett.129.137201}. These exhibit vanishing net magnetization at all temperatures due to antiparallel spin alignment protected by combined symmetry operations, typically time reversal followed by a ninety degrees crystal rotation.
In contrast, full compensation in ferrimagnets arises from oppositely polarized sublattices that are not related by crystal symmetry, and therefore lie outside the altermagnetism classification~\cite{PhysRevX.12.040002}.
Recent important advances have shown that fully compensated ferrimagnetism with altermagnetic-like features can also appear in two-dimensional filling-enforced systems~\cite{2D_FiMs_PRL} and in Luttinger-compensated magnetic phases~\cite{9syc-71w8}. Our approach differs in the fact that full compensation is achieved through an optimal balance of magnetic-moment magnitudes and exchange interaction ratios via targeted chemical substitution.

\section{Discussion and conclusions}

We extend Néel’s classic diagram of ferrimagnetic compensation behavior to explicitly include the effect of magnetic-moment magnitudes, which reveals the stabilization of a ferrimagnetic phase formed by a continuous line of fully compensated points while maintaining strong, eV-scale spin splitting without spin–orbit coupling.
Ferrimagnetic materials tuned to this fully compensated phase are potential candidates to exhibit spin-splitting currents, THz-scale dynamics, and giant or tunneling magnetoresistance. We note that anomalous Hall effects would still require the presence of spin–orbit coupling, as in altermagnets, or a non-collinear magnetic structure such as that reported for Mn$_{4-x}$Z$_x$N ferrimagnets~\cite{ZHANG2022118021, Prendeville_2024}.
Full compensation in ferrimagnets is a fine-tuned cancellation of inequivalent sublattice moments and, therefore, does not generically produce symmetry-protected spin-splitting torques.
Importantly, ferrimagnetic curves closer to the fully compensated line are predicted to show enhanced coercive fields over a broad temperature range.

Our predictions are supported by ab initio- informed Monte Carlo simulations for selected ternary and quaternary GdCo$_5$-type compounds showing a good agreement of the composition-temperature phase diagram with experiment. The extended Néel diagram applies generally to ferrimagnets composed of primary (a) and secondary (b) sublattices with $N_a$ and $N_b$ atoms. Approaching the fully compensated line requires (i) tuning the total local moment toward zero via substitution of weak (e.g., Ni) and strong (e.g., Fe) magnetic atoms while (ii) adjusting the interaction ratio to satisfy $2(\gamma_{aa}-\gamma_{bb})/(-\gamma_{ab})=N_a-N_b$.
We expect that our results will stimulate and guide the search for ferrimagnets tuned towards the fully compensated phase.

\begin{acknowledgements}
We gratefully acknowledge helpful discussions with C.\ E.\ Patrick, J.\ B.\ Staunton, and A.\ Planes.
E.\ M.-T.\ is grateful for financial support from MICIU/AEI/10.13039/501100011033/ and FEDER, UE with Grant. No. PID2024-161052NA-I00, and from CEX2021-001202-M/ AEI/10.13039/501100011033.
The authors acknowledge computing time granted by the supercomputer of the Department of Computational Materials Design, operated by the Max Planck Computing and Data Facility in Garching.

\end{acknowledgements}

\section*{Data Availability}
The data that support the findings of this article are not publicly available. The data are available from the authors upon reasonable request.

\appendix*

\section{Derivation of the Free-Energy Parameters and Compensation-Line Condition}
\label{App1}
\subsection{Relation between real space exchange interactions and free energy magnetic parameters}

The size of $J_{ss'} (\textbf{q})$, introduced in Eq.\ (\ref{eq_FT}), is $n \times n$, where $n$ is the number of atoms inside the primitive cell. Therefore, $n$ = 6 for GdCo$_{5}$-type ferrimagnets. For $\textbf{q}=\textbf{0}$, the matrix can be generally given by eight different parameters:
\begin{equation}
J_{ss'}=
\begin{pmatrix}
\alpha_1 & \beta_1 & \gamma & \gamma & \gamma & \epsilon_1 \\
\beta_1 & \alpha_1 & \gamma & \gamma & \gamma & \epsilon_1 \\
\gamma & \gamma & \alpha_2 & \beta_2 & \beta_2 & \epsilon_2 \\
\gamma & \gamma & \beta_2 & \alpha_2 & \beta_2 & \epsilon_2 \\
\gamma & \gamma & \beta_2 & \beta_2 & \alpha_2 & \epsilon_2 \\
\epsilon_1 & \epsilon_1 & \epsilon_2 & \epsilon_2 & \epsilon_2 & \alpha_3
\end{pmatrix}
,
\label{eq_6x6}
\end{equation}
which reflects the symmetry of the magnetic unit cell. The eigenvector of Eq.\ (\ref{eq_6x6}) with the highest eigenvalue (the most stable magnetic state) that we have found for the materials studied in this work corresponds to a ferrimagnetic structure, i.e., $\textbf{v}_{6\times6}=(m_\text{2c},m_\text{2c},m_\text{3g},m_\text{3g},m_\text{3g},-m_\text{Gd})$, where all the defined quantities are positive. $J_{ss'}$ can be contracted to a $3\times3$ matrix describing the same magnetic state,
\begin{equation}
J_{ss'}^{(3\times3)}=
\begin{pmatrix}
\frac{J_\text{2c-2c}}{2} & \frac{J_\text{2c-3g}}{2} & \frac{J_\text{Gd-2c}}{2} \\
\frac{J_\text{2c-3g}}{3} & \frac{J_\text{3g-3g}}{3} & \frac{J_\text{Gd-3g}}{3} \\
J_\text{Gd-2c} & J_\text{Gd-3g} & J_\text{Gd-Gd}
\end{pmatrix}
.
\label{eq_3x3}
\end{equation}
Note that we have used the same notation as introduced in reference~\cite{chris}.
Table \ref{tab:matrix} gives the relation of the coefficients in matrix $J_{ss'}^{(3\times3)}$ with Eq.\ (\ref{eq_6x6}) such that the eigenvector of the former rightly is $\textbf{v}_{3\times3}=(m_\text{2c},m_\text{3g},-m_\text{Gd})$. An additional contraction to a $2\times2$ matrix can be done,
\begin{equation}
J_{ss'}^{(2\times2)}=
\begin{pmatrix}
J_\text{2c-2c}-\frac{1}{3}J_\text{3g-3g}+\frac{2}{3}J_\text{2c-3g} & J_\text{Gd-3g}-J_\text{Gd-2c} \\
5(J_\text{Gd-3g}-J_\text{Gd-2c}) & J_\text{Gd-Gd}
\end{pmatrix}
,
\label{eq_2x2}
\end{equation}
with the eigenvector associacted to the largest eigenvalue being $\textbf{v}_{2\times2}=(m_\text{TM},-m_\text{Gd})$. Eq.\ (\ref{eq_2x2}) is, however, an approximation for $m_\text{TM}=m_\text{2c}\approx m_\text{3g}$. We can now compare Eq.\ (\ref{eq_G}) and Eq.\ (\ref{eq_2x2}) with $N_\text{TM}=5$. From $-T(\partial S_i/\partial\textbf{m}_i)=\textbf{h}_i$ and
\begin{equation}
\begin{pmatrix}
    h_\text{TM} \\ h_\text{Gd}
\end{pmatrix}
=J_{ss'}^{(2\times2)}\cdot
\begin{pmatrix}
    m_\text{TM} \\ m_\text{Gd}
\end{pmatrix}
,
\label{eq_h_Mdotm}
\end{equation}it directly follows that
\begin{equation}
\begin{split}
   & \gamma_{aa}=\gamma_\text{TM-TM} \approx \frac{J_\text{2c-2c}-\frac{1}{3}J_\text{3g-3g}+\frac{2}{3}J_\text{2c-3g}}{2} \\
   & \gamma_{bb}=\gamma_\text{Gd-Gd} \approx \frac{J_\text{Gd-Gd}}{2} \\
   & \gamma_{ab}=\gamma_\text{TM-Gd} \approx J_\text{Gd-3g}-J_\text{Gd-2c}
   .
\end{split}
\label{eq_JtoU}
\end{equation}
See references~\cite{chris,Edu_Tc} for further detail.

\begin{table}[t]
    \centering
    \begin{tabular}{c|c}
\hline
\textbf{3 $\times$ 3} & \textbf{6 $\times$ 6} \\
\hline
$J_{2c-2c}$ & $(\alpha_1 + \beta_1) \cdot 2$ \\
$J_{3g-3g}$ & $(\alpha_2 + 2\beta_2) \cdot 3$ \\
$J_{\text{Gd-Gd}}$ & $\alpha_3$ \\
$J_{2c-3g}$ & $6\gamma$ \\
$J_{\text{Gd-2c}}$ & $2\epsilon_1$ \\
$J_{\text{Gd-3g}}$ & $3\epsilon_2$ \\
\hline
    \end{tabular}
\caption{Relation between coefficients of matrices given in Eq.\ (\ref{eq_3x3}) and Eq.\ (\ref{eq_2x2}).}
\label{tab:matrix}
\end{table}

\subsection{Condition for the fully compensated phase}

Using Eq.\ (\ref{eq_JtoU}), Eq.\ (\ref{eq_2x2}) can be rewritten in terms of the Gibbs free energy coefficients as
\begin{equation}
J_{ss'}^{(2\times2)}=
\begin{pmatrix}
2\gamma_\text{TM-TM} & N_\text{Gd}\gamma_\text{TM-Gd}  \\
N_\text{TM}\gamma_\text{Tm-Gd} & 2\gamma_\text{Gd-Gd}
\end{pmatrix}
,
\label{eq_2x2gammas}
\end{equation}
where we have reinstated $N_\text{TM}$ and $N_\text{Gd}$ for generality. The total net magnetization vanishes at all temperatures at the fully compensated line. At 0 K this condition requires $N_\text{TM}\mu_\text{TM}=N_\text{Gd}\mu_\text{Gd}$, whereas in the paramagnetic limit ($T\rightarrow T_\text{tr}$) it imposes $N_\text{TM}\mu_\text{TM}m_\text{TM}+N_\text{Gd}\mu_\text{Gd}m_\text{Gd}=0$. Combining these two relations with the fact that $(m_\text{TM}, m_\text{Gd})$ is the eigenvector of Eq.\ (\ref{eq_2x2gammas}), we obtain $m_\text{TM}=-m_\text{Gd}$ and the following condition for the fully compensated phase
\begin{equation}
    \frac{2(\gamma_\text{TM-TM}-\gamma_\text{Gd-Gd})}{-\gamma_\text{TM-Gd}}=N_\text{TM}-N_\text{Gd}
    .
    \label{eq_cond_app}
\end{equation}

\bibliography{./bibliography.bib}
\end{document}